# Intrinsic-to-extrinsic transition in fracture toughness through structural design: A lesson from nature

Bin Liu[1], Yanjie Jia[1], He-Ling Wang[1], Huajian Gao[2]

[1]*Department of Engineering Mechanics, Center for Mechanics and Materials, Tsinghua University, Beijing 100084, China*

Email address: liubin@tsinghua.edu.cn

[2]*School of Engineering, Brown University, Providence, Rhode Island 02912, USA*

Email address: Huajian_Gao@brown.edu

## Abstract

Catastrophic failure of materials and structures due to unstable crack growth could be prevented if facture toughness could be enhanced at will through structural design, but how can this be possible if fracture toughness is a material constant related to energy dissipation in the vicinity of a propagating crack tip. Here we draw inspiration from the deformation behavior of biomolecules in load bearing biological materials, which have been evolved with a large extensibility and a high breaking strength beyond their elastic limit, and introduce an effective biomimetic strategy to enhance fracture toughness of a structure through an intrinsic to extrinsic (ITE) transition. In the ITE transition, toughness starts as an intrinsic parameter at the basic material level, but by designing a protein-like effective stress-strain behavior the toughness at the system level becomes an extrinsic parameter that increases with the system size without bound. This phenomenon is demonstrated through a combination of numerical simulations, analytic modeling and experiments, and leads to a biomimetic strategy which can be broadly adopted to enhance fracture toughness in engineering systems.

**Significance Statement**

Motivated by the deformation behavior of biomolecules in load bearing biological materials that have been evolved with a large extensibility and a high breaking strength beyond their elastic limit, here we introduce a biomimetic strategy to enhance facture toughness at the system level through an intrinsic to extrinsic (ITE) transition. We demonstrate that this phenomenon is based on well-established principles in mechanics and can be broadly applied in engineering applications.

**Main Text**

Load bearing biological materials such as bone (1-3) exhibit interesting mechanical properties such as superior strength and toughness (4-8). Some of the toughest sea shells (9, 10), which consist of mineral plates embedded in protein-rich organic materials, have been reported to have fracture energy 3 orders of magnitude higher than that of the mineral (4, 5); dragline spider silk and other natural fibers show breakage energy two orders of magnitude greater than that of high strength steel (6). The superior properties of biological materials have been attributed to the parallel staggered arrangement of mineral plates in the biological nanostructure (7, 11-17) as well as the special force-extension curves of biomolecules that constitute their organic phase (18, 19). For further surveys on theoretical and experimental studies in the literature on mechanical principles of biological materials including recent developments in biomimetic materials, the reader can be referred to some recent review articles (20, 21). These studies are also calling for further efforts to develop general biomimetic strategies that can be broadly applied in engineering systems.

Here we draw inspiration from the typical deformation behavior of biomolecules in load bearing biological materials which have been evolved with a large extensibility and a high breaking strength beyond their elastic limit. It will be shown that such behavior, achieved through the hierarchical structure of biomolecules rather than the chemical bonds in their primary structure, can lead to an intrinsic to extrinsic (ITE) transition in fracture toughness of the material. Through the ITE transition, fracture toughness is no longer a material constant, rather it becomes an extrinsic parameter that increases with system size. We will show that this strategy can be widely adopted to enhance the fracture toughness at the system level.

Figures 1(a,b) show a typical three-stage force-extension behavior of biomolecules that are thought to enhance the fracture toughness of biological materials (18, 19), namely stage I — elastic stretch, which is recoverable upon unloading; stage II — domain unfolding, which is characterized by a sawtooth-like behavior with extraordinary extensibility (22); and stage III — backbone stretching with a significant increase in force. While fracture in elastic-plastic engineering materials has been extensively studied in the past (23-30), the deformation behavior of biomolecules deviates from that of engineering materials in a number of important aspects, including 1) stage II giving a large extensibility and 2) stage III leading to a high ultimate tensile strength beyond the elastic limit at the end of stage I.

We hypothesize that the deformation behavior of biomolecules in load bearing biological materials is a key for these materials to achieve their superior mechanical properties including an amazing resistance to catastrophic, unstable crack propagation. To test this hypothesis, consider a strip of material consisting of a triangular network of molecular chains, as shown in Fig. S1 in the SI; each chain follows the characteristic force-extension relation of a protein as shown in Fig. 1(b). The strip contains a pre-crack along the center line, with upper and lower boundaries subjected to increasing displacements in the normal direction. A single molecular chain would fracture once its ultimate tensile strength is reached, leading to crack propagation. Such a network of molecular chains has been used to study dynamic crack propagation (31, 32, 33) and fracture behavior of proteins (34). As stretching on the strip boundary increases, the material surrounding the crack tip undergoes inelastic deformation, and the inelastic zone enlarges and saturates (see Fig. S1(c)), leading to crack growth and steady-state crack propagation. The steady state fracture energy $G_{IC}$ (energy dissipated per unit area of crack propagation) is obtained from the difference in energy stored far ahead and far behind the crack tip, and the fracture toughness $K_{IC}$ is related to the fracture energy $G_{IC}$ by $G_{IC} = K_{IC}^2 / E$ (35), where $E$ is Young's modulus of the strip.

Our analysis indicates that the sawtooth feature of the force-extension curve of a biomolecule has an insignificant effect on the fracture toughness of the network (see SI and Fig. S2), so that Stage II can be approximated by a straight segment characterized by a maximum inelastic strain $\varepsilon_T$ [Fig. 1(b)]. For simplicity, the stage-III slope is taken to be the same as that for stage I. Figure 1(d) shows the distribution of inelastic strain in the strip for quasi-static, steady-state crack propagation for several ratios of $\varepsilon_T/\varepsilon_Y$, where $\varepsilon_Y$ is the yield strain (i.e., elastic limit) [Fig. 1(b)]. The figure for $\varepsilon_T = 0$ shows no inelastic deformation since it corresponds to stage-I elastic deformation followed immediately by chain fracture, and the corresponding fracture energy and fracture toughness are denoted by $G_{IC}^{(0)}$ and $K_{IC}^{(0)}$, respectively. The inelastic zone clearly increases with $\varepsilon_T$, leading to a rapidly increasing fracture energy $G_{IC}$, as shown in Fig. 1(e). For example, $G_{IC}$ for $\varepsilon_T/\varepsilon_Y = 3$ is 27 times of $G_{IC}^{(0)}$ for $\varepsilon_T = 0$. Figure 1(e) contains results for different ratios of ultimate tensile strength to yield stress, $\sigma_{UTS}/\sigma_Y = 1$, 1.5 and 2 (illustrated in Fig. S1(b)), which are essentially the same.

The above relation between $G_{IC}$ and $\varepsilon_T/\varepsilon_Y$ can also be obtained analytically. Figure 1(c) shows a schematic diagram of inelastic zone, where the same colors as in Fig. 1(b) are used to denote regions corresponding to stages I, II and III. The inelastic strain relaxes the stress level around the crack tip and reduces the crack-tip stress intensity factor, $K_{tip}$, by

$$K_{tip} = K_{applied} - \Delta K, \qquad (1)$$

where $K_{applied}$ is the applied stress intensity factor of the remote field [Fig. 1(c)], $\Delta K (>0)$ is the net contribution from the inelastic strain and is given by (see SI)

$$\Delta K = AE\varepsilon_T\sqrt{2\pi w_Y}, \qquad (2)$$

where $A$ is a non-dimensional material constant on the order of 1, and the height $w_Y$ of inelastic zone [Fig. 1(c)] is defined as $w_Y = \left(K_{applied}/\sigma_Y\right)^2/(2\pi)$ following linear elastic fracture mechanics (29). For steady-state crack propagation, $K_{tip} = K_{IC}^{(0)} = \sqrt{EG_{IC}^{(0)}}$ and the corresponding $K_{applied}$ is the fracture toughness $K_{IC}$. Equations (1) and (2) can be combined to give the following relation

$$\frac{K_{IC}^{(0)}}{K_{IC}} + A\frac{\varepsilon_T}{\varepsilon_Y} = 1 \qquad (3)$$

for fracture toughness and

$$\frac{G_{IC}}{G_{IC}^{(0)}} = \left(1 - A\frac{\varepsilon_T}{\varepsilon_Y}\right)^{-2} \qquad (4)$$

for fracture energy. Figure 1(e) shows that the above equation agrees well with the simulation results based on the triangular network of molecular chains.

Equation (4) suggests a critical ratio, $\varepsilon_T/\varepsilon_Y = A^{-1}$, at which the steady-state fracture energy becomes infinite. Note that the fracture energy of a finite sized sample is always finite. In the strip model under consideration, the infinite fracture energy is a result of taking the strip length to be infinite. In practical terms, this means that the plastic zone spreads over to the entire sample and is no longer confined to the crack tip region, so that the fracture energy is no longer an intrinsic parameter, rather it becomes an extrinsic parameter that scales with the overall size of the sample. This is intuitively referred to as an intrinsic-to-extrinsic transition, or simply an ITE transition, in fracture toughness. The critical ratio for ITE transition is found to be ~3.5 for the strip with the triangular network of chains in Fig. S1(a). For $\varepsilon_T/\varepsilon_Y$ exceeding this critical ratio, crack propagation never attains a true steady state, as shown in Fig. 1(d) for $\varepsilon_T/\varepsilon_Y = 4$. The inelastic zone continues to grow and reaches the boundary of the sample, thereby de-localizing energy dissipation associated with crack growth, as to be further discussed in experiments in Fig. 3.

The linear relation between the reciprocal of fracture toughness $1/K_{IC}$ and the ratio $\varepsilon_T/\varepsilon_Y$ in Eq. (3), shown in Fig. 1(f), is in excellent agreement with our simulation results. The intercept of this curve with the horizontal axis corresponds to the critical ratio $\varepsilon_T/\varepsilon_Y$ for ITE transition.

Even though the ratio of ultimate tensile strength to yield stress $\sigma_{UTS}/\sigma_Y$ does not affect the quasi-static fracture toughness [Figs. 1(e), (f)], it plays a critical role in dynamic crack propagation. Figure 2(a) shows two force-extension curves with the same ultimate tensile strength $\sigma_{UTS}$ and same ratio $\varepsilon_T/\varepsilon_Y = 1.5$, but different yield stresses. At the applied strain of 0.0045, both cases attain quasi-static steady-state crack propagation with the same fracture toughness, but the inelastic strain for the case of relatively low yield stress $\sigma_Y = 0.5\sigma_{UTS}$ is much more de-localized than that for relatively high yield stress $\sigma_Y = \sigma_{UTS}$, as shown in Fig. 2(b). Figure 2(c) gives snapshots for the inelastic strain distribution at the applied strain rate of $2.5 \times 10^{-5} c_l/l_0$ [up to the same applied strain as in Fig. 2(b)], where $c_l$ is the longitudinal wave speed and $l_0$ is the chain length (see SI). The crack propagates continuously through the entire specimen for the high yield stress case, but it propagates and then stops for the low yield stress case (see movie1 in the SI). The reason for this difference can be attributed to the fact that stress fluctuations during dynamic crack propagation force more materials into inelastic deformation at lower yield stress.

To validate the theoretical model described above, fracture experiments were conducted for polyvinylchloride (PVC) and polyethylene (PE) strips in Figs. 3(a)-(b), respectively. These two materials have very different force-extension curves: PVC behaves elastic without significant inelastic deformation, while PE exhibits large ductility with relatively small yield stress compared to its ultimate tensile strength, as shown in the first row. The second row of Figs. 3 shows the distributions of inelastic strain calculated from the computational model in Fig. S1(a) adopting the corresponding force-extension curves. In contrast to the case of PVC, the crack tip in PE is significantly blunted and the inelastic deformation spreads all the way to the boundary, i.e., extrinsic fracture toughness. These observations are consistent with experiments shown in the last two rows of Figs. 3. For PVC, the crack propagated unstably across the specimen. In contrast, the crack in PE propagated stably and slowly after an initial time interval of $\Delta t = 100s$, as loading rises continuously (see movie2 in the SI). Substantial differences were also seen on pre-marked grids on the specimens after crack propagation, which were essentially unchanged near the top and bottom boundaries in PVC but severely stretched in PE, suggesting that inelastic deformation reached the boundaries in the latter case even for specimen size as large as 180cm x 40cm.

The extrinsic fracture toughness in Fig. 3(b) can be tuned to become intrinsic by modifying the force-extension curve via pre-straining. Figure 3(c) shows the results for a pre-strained PE sample with approximately the same ultimate tensile strength as that shown in Fig. 3(b), but much smaller ductility. In this case, simulations show that the inelastic deformation does not spread to the boundaries, which is supported by the experimental observation that the grids were essentially unchanged near the top and bottom boundaries, and that crack propagation was clearly unstable with rapid growth over a short time interval $\Delta t = 1s$ (see movie3 in the SI).

The intrinsic fracture toughness can also become extrinsic. Consider a flat aluminum strip which has a yield stress approximately the same as its ultimate tensile strength and relatively low ductility, as shown in the first row of Fig. 4(a). These properties should lead to intrinsic fracture toughness and therefore unstable crack growth, as confirmed by experiments shown in the last two rows of Fig. 4(a) before and after crack propagation (also see movie4 in the SI). To achieve extrinsic fracture toughness and therefore prevent unstable crack growth, we fabricated the aluminum strip into a "battlement" shape, inspired by the domains in protein, to achieve large ductility while reducing effective yield stress for such a structure, as shown in the second row of Fig. 4. Comparison of the last two rows of Fig. 4(b) indicates that the battlement shape ahead of the propagating crack tip was almost completely flattened out upon loading, suggesting that inelastic deformation reached out to the boundaries, stabilizing crack growth (see movie5 in the SI).

In conclusion, we have identified a general, bio-inspired strategy to fabricate fracture resistant materials and structures. The central point of this strategy is to design a protein-like stress-strain curve such that energy dissipation ahead of the crack tip is not confined to the vicinity of the crack tip, but rather spread to the entire sample so that the fracture toughness is no longer an intrinsic material property but an extrinsic quantity depending on the specimen size. Such a strategy should be generally useful for designing engineering materials and structures against catastrophic failure due to unstable crack growth. For example, the cellular materials, including lattice materials with regular holes and metal foams with irregular pores, can achieve the desired force-extension curve by adjusting microstructures to realize ITE transition of fracture toughness. By engineering nanoscale twin boundaries in metal materials, the stage II of the stress-strain curve can be substantially elongated (36, 37), which could enhance energy dissipation near the crack tip and make possible the fracture tolerant behavior.

**Acknowledgements**

The authors acknowledge the support from National Natural Science Foundation of China (Grant Nos. 11425208, 11372158, and 51232004), and Tsinghua University Initiative Scientific Research Program (No. 2011Z02173). Our special thanks are due to Dr. Wan-Ting Ren for her constructive and effective suggestions in conducting this study.

# Figure Captions

**Figure 1.** Computational model of crack propagation in a strip made of protein-like materials.
(a) Three-stage deformation of a protein molecule under stretch: the elastic deformation with crosslink network untouched (stage I), domain unfolding (stage II) and stretching of backbone (stage III).

(b) Typical force-extension curve for a protein molecule, with essential features showing that 1) stage II (and III) gives large inelastic deformation (i.e., large ductility), and 2) the ultimate tensile strength at the end of stage III is much larger than the elastic limit at the end of stage I.
(c) Inelastic zones around a steady-state crack, colored after deformation regions corresponding to stages I, II and III.
(d) Maps of inelastic strain associated with quasi-static, steady-state crack propagation for several ratios of $\varepsilon_T/\varepsilon_Y$, showing increasing inelastic zone with $\varepsilon_T$.
(e) The calculated fracture energy $G_{IC}$ for different ratios of ultimate tensile strength to yield stress, $\sigma_{UTS}/\sigma_Y =1$, 1.5 and 2 (illustrated in Fig. S1(b)). Note that $G_{IC}$ increases rapidly with $\varepsilon_T/\varepsilon_Y$.
(f) Predicted linear relation between the reciprocal of fracture toughness $1/K_{IC}$ and the strain ratio $\varepsilon_T/\varepsilon_Y$ in Eq. (3), with excellent agreement with numerical simulation results.

**Figure 2.** Simulations of dynamic crack propagation in a strip made of a network of chain molecules with different force-extension curves.
(a) Two force-extension curves with the same ultimate tensile strength $\sigma_{UTS}$ and ratio $\varepsilon_T/\varepsilon_Y =1.5$, but different yield stress.
(b) At the applied strain of 0.0045, the two systems with force-extension curves shown in Fig. 2(a) reach quasi-static steady-state crack propagation with identical fracture toughness, but inelastic strain in the case of lower yield stress $\sigma_Y = 0.5\sigma_{UTS}$ is much more de-localized than that for higher yield stress $\sigma_Y = \sigma_{UTS}$.
(c) Snapshots of inelastic strain distributions during dynamic crack propagation under applied strain rate of $2.5\times10^{-5}\, c_l/l_0$ [up to the same applied strain as Fig. 2(b)]. The results show that the crack propagates continuously through the entire specimen at higher yield stress, but it propagates a while and then stops at lower yield stress (see movie1 in the SI).

**Figure 3**.
   Experiments on crack propagation in polyvinylchloride (PVC) and polyethylene (PE) strips. The first row in (a)-(b) shows that the force-extension curve is mainly elastic with little inelastic deformation for PVC, while plastic with large ductility and a relatively small yield stress compared to the ultimate tensile strength for PE. The distributions of inelastic strain calculated based on the computational model in Fig. S1(a) adopting these curves are shown in the second row. In contrast to the PVC case, the crack tip in the PE strip is significantly blunted with inelastic deformation spreading all the way to the boundaries. The model predictions are corroborated with corresponding experimental observations shown in the last two rows. In the case of PVC, fracture occurred via instantaneous, unstable crack propagation across the whole specimen. The crack in PE, however, propagated stably and slowly as increasing displacement is imposed on the boundary. There was essentially no crack growth over a time interval of $\Delta t = 100s$ (see movie2 in the SI).

Figure 3(c) shows the results for an identical but pre-strained PE sample. The first row shows that the pre-strained sample has approximately the same ultimate tensile strength as the unstrained sample in Fig. 3(b), but with much smaller ductility. Compared to Fig. 3(b), the inelastic deformation obtained from the computational model for the pre-strained force-extension curve does not spread to the boundaries. This is consistent with experiments with the pre-strained PE samples showing that the pre-marked grids were essentially unchanged near the top and bottom boundaries, and that crack propagation was clearly unstable, with rapid growth over a short time interval of $\Delta t = 1s$ (see movie3 in the SI).

**Figure 4.** Controlling fracture resistance of materials by tuning force-extension curve.
(a) The first row shows a flat aluminum strip and its force-extension curve, which has the yield stress approximately the same as the ultimate tensile strength with relatively low ductility. The last two rows show the experimental results before and after crack propagation, indicating unstable crack growth in the flat aluminum strip (see movie4 in the SI).
(b) The first row gives the tensile force-extension curve of a "battlement" shaped aluminum strip, with large ductility and relatively low effective yield stress. Comparison of the last two rows indicates that the battlement shaped structure ahead of the crack tip was almost completely flattened out upon loading, suggesting that inelastic deformation reached out to the boundaries, stabilizing crack growth (see movie5 in the SI).

**Figure 1.**

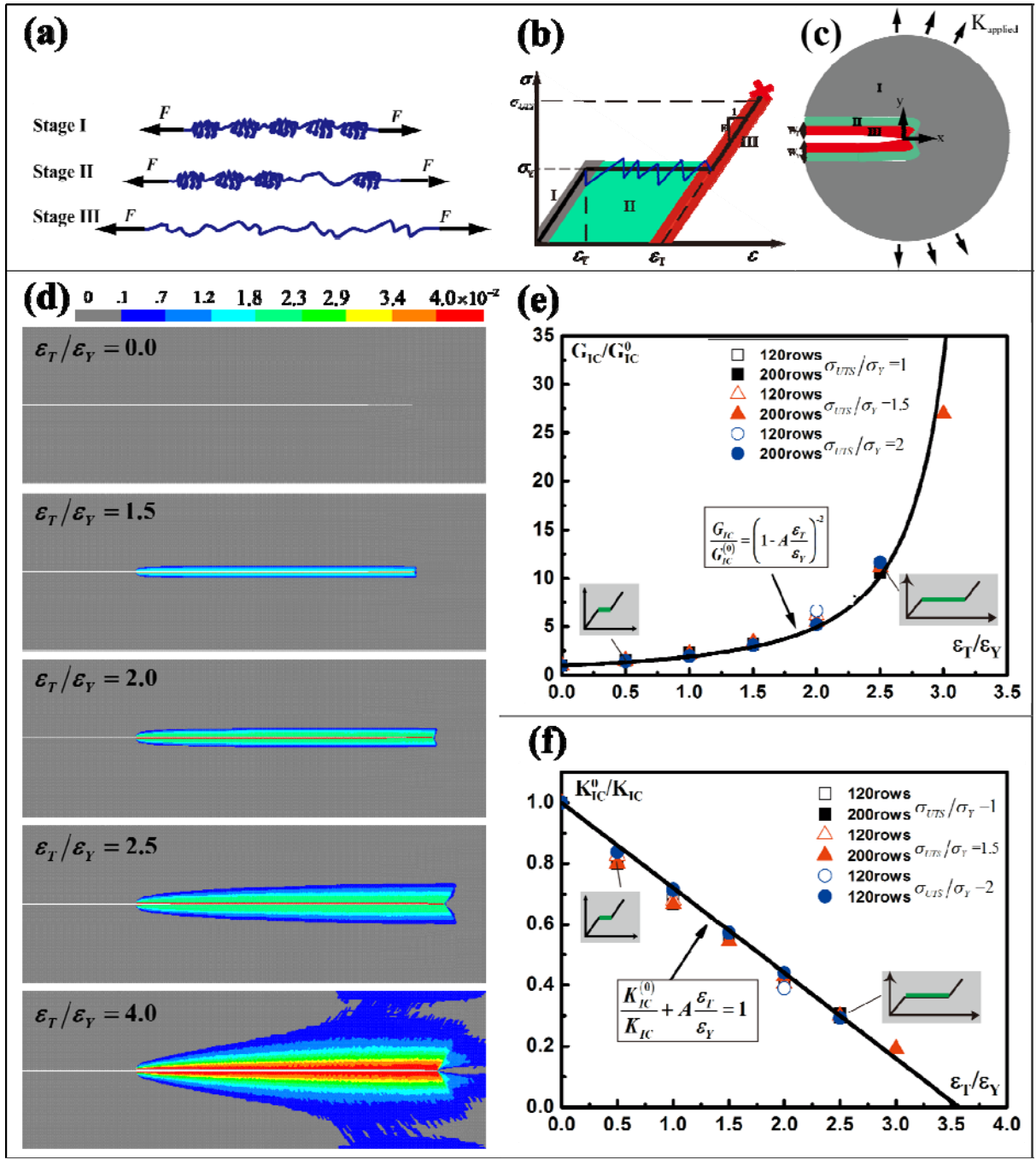

**Figure 2**

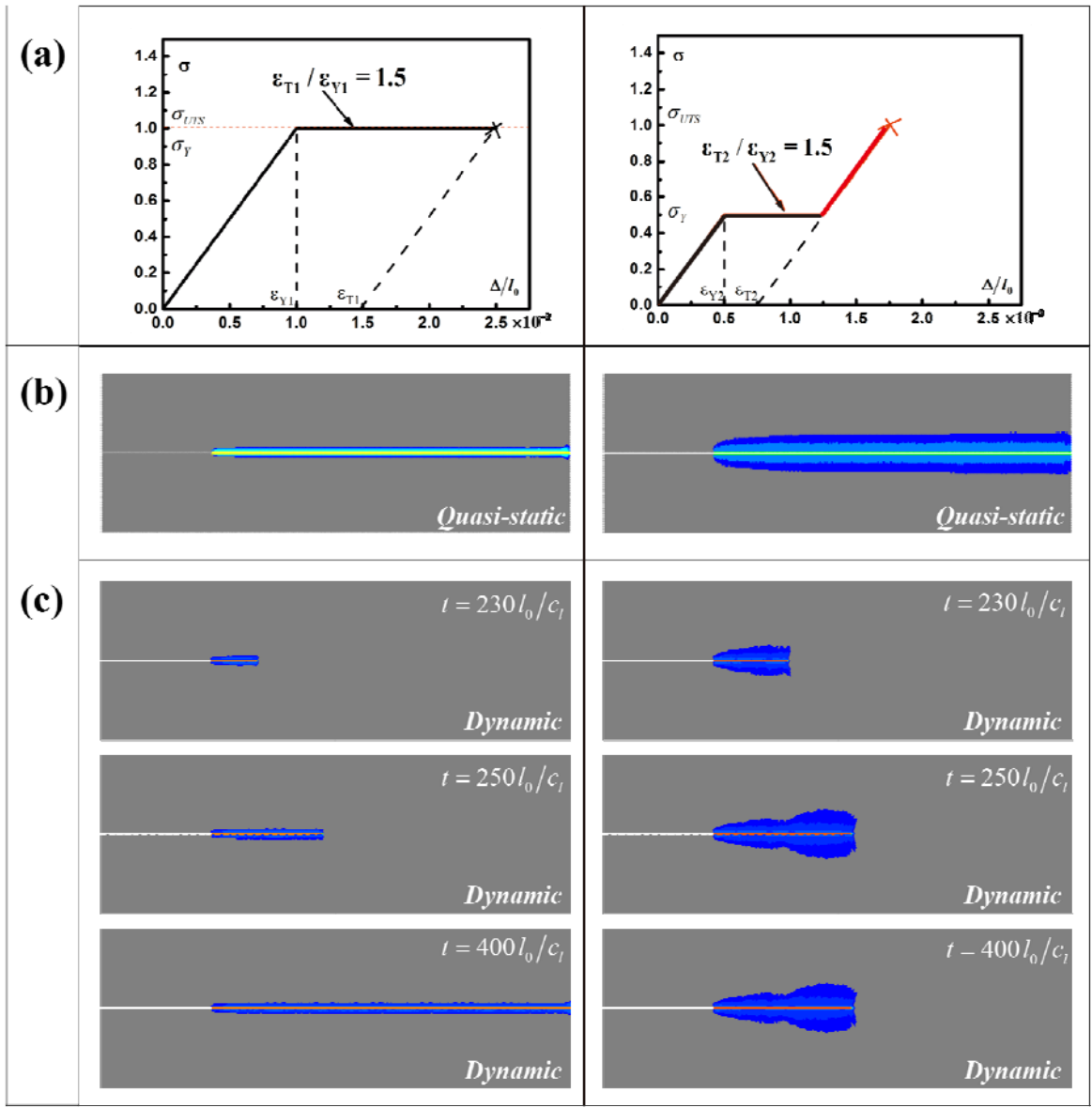

**Figure 3**.

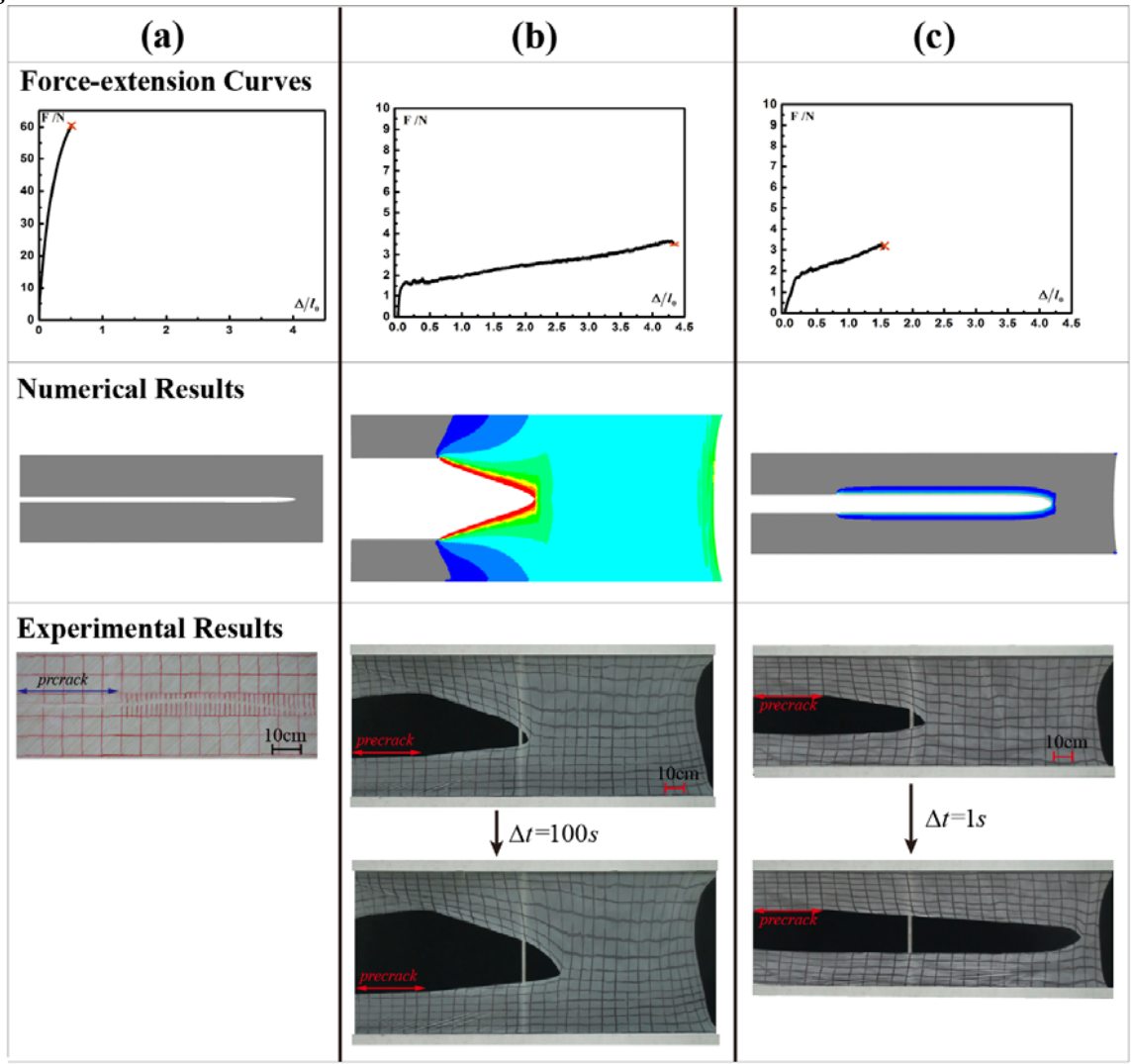

**Figure 4.**

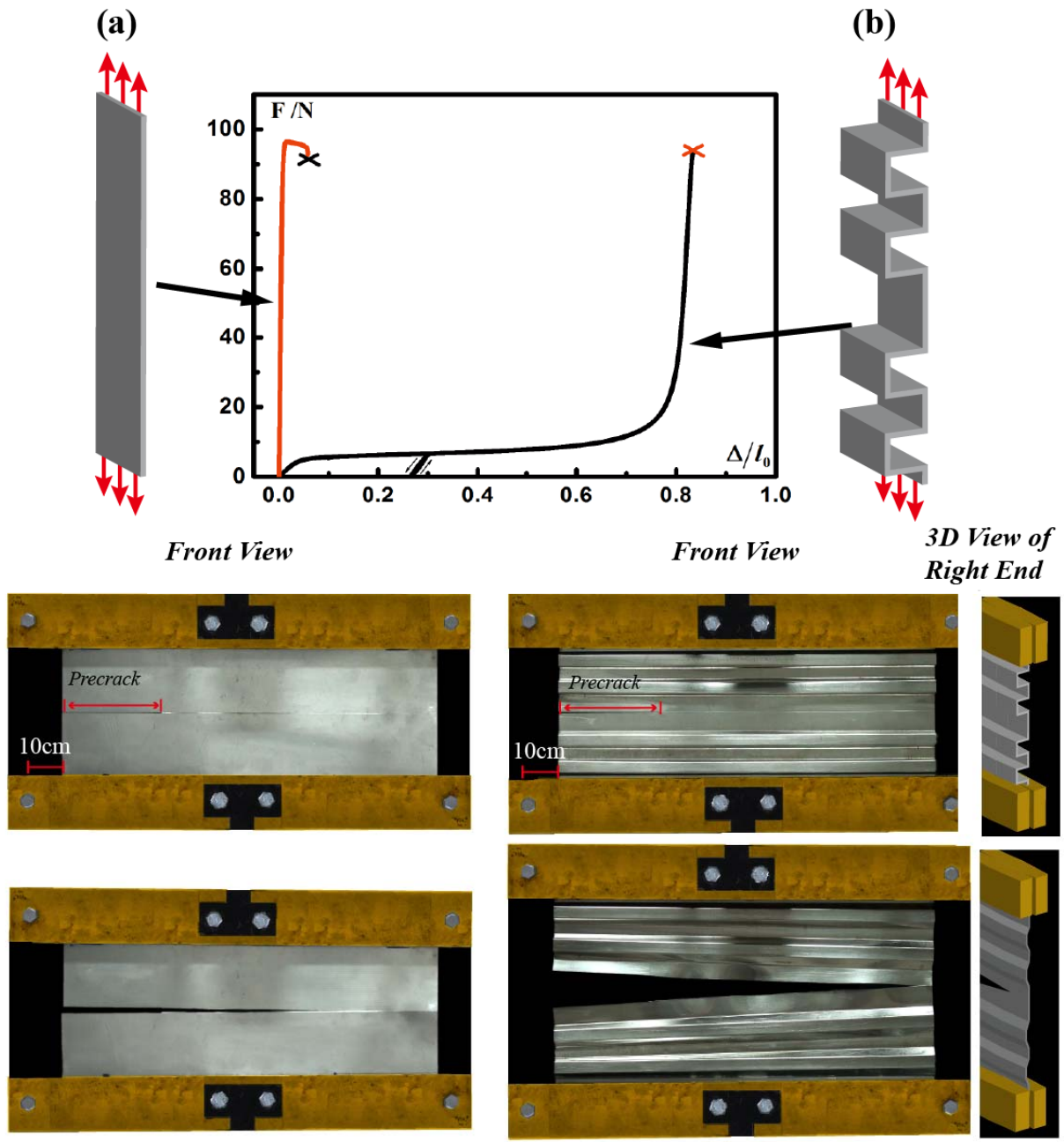